\documentstyle[12pt]{article}

\def\be{\begin{equation}}
\def\ee{\end{equation}}
\def\bea{\begin{eqnarray}}
\def\eea{\end{eqnarray}}
\def\d{\partial}
\def\D{\nabla}
\def\scv{-\sum_{s=1}^{m}}
\def\sct{\sum_{r=1}^{l}}
\def\icv{\nu_1,...,\nu_m }
\def\ict{\mu_1,...,\mu_l}
\def\icvv{\nu_1,.,\sigma,.,\nu_m}
\def\ictt{\mu_1,.,\sigma,.\mu_l}
\def\iccv{^{\sigma}_{\rho\, \nu_s}}
\def\icct{^{\mu_r}_{\rho\,\sigma}}
\def\cvV{V^{\ict}_{\icvv}}
\def\ctV{V^{\ictt}_{\icv}}
\def\ic{^{\rho}_{\mu\nu}}
\def\t{\tilde}
\def\g{\gamma}
\def\G{G_{00}}

\begin{document}
\vspace*{2cm}
\begin{center}
{\large\bf GEOMETRY OF T-DUALITY}
\vskip 1.5 cm
{\bf  Javier Borlaf}\footnote{\tt javier@delta.ft.uam.es}
\vskip 0.05cm
Departamento de F\'{\i}sica Te\'orica, Universidad Aut\'onoma,
28049 Madrid, Spain
\end{center}

\setcounter{page}{0} \pagestyle{empty}
\thispagestyle{empty}
\vskip 2cm
\begin{abstract}
A "reduced" differential geometry adapted to the presence of
abelian isometries is constructed. Classical T-duality diagonalizes
in this setting, allowing us to get conveniently the transformation
of the relevant geometrical objects such as connections, pullbacks 
and generalized curvatures. Moreover we can induce privileged 
maps from the viewpoint of covariant derivatives in the target-space
and in the world-sheet generalizing previous results, 
at the same time that we can correct connections 
and curvatures covariantly in order to have a proper transformation under 
T-duality.

\end{abstract}

\vfill
\begin{flushleft}
FTUAM-97-8 \\
hep-th/9707007\\
June 1997
\end{flushleft}
\newpage\pagestyle{plain}

\section{Introduction}
T-duality is a fundamental tool in the understanding of  the 
fashionable string dualities \cite{PCJ}. 
The elementary statement of the T-duality establishes that the 
perturbative spectrum of a string theory with a dimension
compactified on a circle of radius R, is equivalent to the
one compactified in a circle of radius $1/R$, provided
we interchange winding and momentum quantum numbers at 
the same time that we transform the string  coupling
constant \cite{GPR,AO}.
If we allow the presence of a generic geometry 
(metric $G_{\mu\nu}$
, torsion potential $B_{\mu\nu}$ and dilaton $\Phi$) 
having an abelian Killing vector in the compactified 
direction $X^0$,
the backgrounds resulting to be 1-loop (conformally)equivalent 
are given by the Buscher's
formulas \cite{BU}:

\bea
\label{BF}
&\t G_{00}= 1/\G \nonumber\\
&\t G_{0i}= B_{0i}/\G\,\,\,\,\,\,\
\t B_{0i}= G_{0i}/\G\nonumber\\
&\t G_{ij}= G_{ij} - (G_{0i}G_{0j}-B_{0i}B_{0j})/
\G\nonumber\\
&\t B_{ij}= B_{ij} - (G_{0i}B_{0j}-B_{0i}G_{0j})/
\G\nonumber\\
&\t \Phi = \Phi - \frac{1}{2}\ln \G
\eea

In recent years a non-perturbative  usage of T-duality
has been made in the context of open strings playing
with the map of the Neumann-Dirichlet boundary conditions.
The promotion of 
the hypersurfaces in which strings rest their endpoints 
to be dynamical extended objects called Dirichlet-branes,
allow their identification with the carriers of  the RR-charges 
required by the string duality at the same time that it makes 
doubtful the name of "string theory" for the resulting scenario
\cite{DLP,JKKM}.

Many features of this 
topic of T-duality have been developed extensively in
the literature
(\cite{OA,K,H,GPR}).
Despite that important effort, it seems to be a 
lack of a systematic study of the mapping between 
geometries for the stringy  (1-loop) equivalent space-times. 
This gap is related with the nature of the non-linear map
(\ref{BF}) which highly complicates the calculations
for the transformation of geometrical objects such as the
generalized connection and its curvature, privileged maps
for the covariant derivatives and pullbacks, and many others.

In order to clarify all these points it is presented a sort of 
"parallel differential geometry" that we will call "Reduced
Geometry", having the property of being adapted to the
presence of abelian Killing vectors (in fact it is only defined 
in that context). We will see how T-duality transformations 
diagonalize in this setting for the main geometrical objects, 
including the generalized curvature ; for the later, we found 
for the first time its complete transformation, which can be 
expressed 
in a covariant way in terms of itself and of the Killing 
vector.

Moreover, a "canonical" T-duality transformation 
is constructed for arbitrary tensors with the property 
of  transforming linearly
the covariant derivatives calculated from the generalized 
connections.
These results unify and generalize the map obtained for the
p-forms in \cite{K}, and it includes the fundamental one of the
complex structures for holomorphic Killing vectors \cite{H}.

We can extend the result giving above to the "canonical" 
T-duality 
transformation of maps from the world-sheet to the tangent
space of our target-space manifold. The clasical string
dynamics will be the most representative example of this
"canonical" map.

In section 2 we define and describe the construction 
of the "reduced
geometry" giving the basic map relating "usual" and "reduced"
objects (generic tensors, connections and curvatures).
In section 3 we get the "canonical" T-duality map, relating
linearly covariant derivatives, and a "non-canonical"
one relating linearly "covariant divergences".
In section 4 we obtain the "canonical" map for the 
classical world-sheet dynamics. Section 5 shows the generalized
curvatures' transformation and the minimal correction for them
to transform linearly under T-duality. As a straightforward
outcome I rederive
the 1-loop beta function's transformation. In section 6 we found 
a "canonical" covariant derivative commuting with the "canonical"
T-duality transformation. It is used to get a set 
of new T-duality scalars. 
In the Apendix we summarize the basic formulas.


\section{Reduced Differential Geometry}
In this section  will be built a parallel tensor calculus 
for manifolds with
an structure endowed with abelian Killing vectors. 
The main objective is to
exploit the presence of these Killings in order to get a 
strongly
simplified structure that I will call the Reduced Geometry. 
As we will see,
that structure is nicely adapted to T-duality.

Let us assume the existence of a set of n commuting vector fields
$\{K^{\mu}_{a}\} $ with $\{\mu,\nu =0,1,..., D-1\}$, 
$\{a,b=1,...,n\}$ and
D the dimension of the manifold M.

We restrict our attention to the  space  $\Omega$ of tensors 
V  in M satisfying
\be
\label{lie}
{\cal L}_{k_a} V = 0
\ee
that means simply that we can choose  coordinates 
$\{x^i,\, x^a \}$ with
$i=n,...,D-1$ ,called adapted coordinates, in which V does not
depend on $x^a$,
ie., $\d_{a} V = 0 $.

The covariant differentiation is not a mapping in $\Omega$, 
or in other words,
its commutator with the Lie
derivative is in general non vanishing :
\be
\label{lie2}
[{\cal L}_{k_a} \, , \D_{\rho}] V^{\ict}_{\icv} = \scv ({\cal
L}_{k_a}\D)\iccv \cvV \,
+ \, \sct ({\cal L}_{k_a}\D)\icct \ctV
\ee
where I have defined \footnote{the conventions for the
curvature can be found in the Apendix}
\be
\label{liec}
({\cal L}_{k_a} \D)^{\sigma}_{\mu\,\nu} \equiv 
K^{\alpha}_{a}R_{\alpha\mu\nu}^{\sigma} \, + 
\D_{\mu}\D_{\nu}K^{\sigma}_{a}
\, + 2\D_{\mu}(K^{\alpha}_{a}T_{\alpha\,\nu}^{\sigma})
\ee
being $R^{\sigma}_{\alpha\mu\nu}$ the curvature for the 
connection 
$\Gamma_{\lambda\beta}^{\delta}$ and $T\ic$ the corresponding 
torsion. It 
can be checked that (\ref{liec}) reduces to the 
desired $\d_{a}\Gamma^{\sigma}_{\mu\,\nu} $ in adapted 
coordinates.

If we are interested in connections preserving the condition 
(\ref{lie}), we must impose  
\be
\label{liec0}
{\cal L}_{k_a}\D = 0
\ee

Then, I have established our framework throught the conditions 
(\ref{lie}) and (\ref{liec0}).Moreover I assume the choice of 
adapted 
coordinates to the Killings.There is a freedom for that choice 
that is 
reflected in the existence of a subset of diffeomorphisms 
(adapted 
diffeomorphisms) relating the different possibilities.
Modulo arbitrary changes in the $x^i$ 
transverse coordinates, the relevant adapted ones are 
\bea
\label{ad}
&x^{'i} = x^i \nonumber\\
&x^{'a} = x^{a} + \Lambda^a (x^j) 
\eea
Tensors in  $\Omega$ transforms linearly under this change as 
is expected
\be
\label{tt1}
V(x^i)^{' \ict}_{\icv} = J(\d \Lambda)^{\mu_1 ,..,\mu _l  
;\alpha_1
,..,\alpha_m
}_{\nu_1 ,..,\nu_m ;\beta_1, ..,\beta_l}
V(x^j)^{\beta_1 ,...,\beta_l }_{\alpha_1 ,...,\alpha_m }
\ee
 where I define
\be
\label{j}
J(\d \Lambda)^{\mu_1 ,..,\mu _l  
;\alpha_1 ,..,\alpha_m }_{\nu_1 ,..,\nu_m
;\beta_1, ..,\beta_l}
\equiv \prod^{l}_{r=1}J^{\mu_r }_{\beta_r }(\d \Lambda)\,
\prod^{m}_{s=1}J^{\alpha_s }_{\nu_s }(-\d \Lambda)
\ee
 
\be
\label{j2}
J^{\mu }_{\nu}(\d \Lambda) = \delta^{\mu }_{\nu} +  
\delta^{\mu }_{a}\d_{i}\Lambda^{a} \delta^{i }_{\nu}
\ee

Using a short notation

\be
\label{tt}
V' = J(\d \Lambda)V
\ee
Because the abelian nature of the 
diffeomorphism, J provides a representation of  $U(1)^n$ in the 
vector space of tensors of the same rank than V satisfying 
(\ref{lie}); in 
particular, $J(\d\Lambda_1)J(\d\Lambda_2)=
J(\d(\Lambda_1 + \Lambda_2))$ and
$J(0)={\bf 1}$ implying $J^{-1}(\d\Lambda)=J(-\d\Lambda)$.  

In the cases we are interested, it is natural to find a set of 
"transverse" gauge fields $\{A^{a}_i(x^j)\}$ transforming under 
the adapted diffeomorphism as
\be
\label{abcdexx}
A'^{a}_{i}(x^j)=A^{a}_{i}(x^j) - \d_{i}\Lambda^{a}(x^j)
\ee
Now I can define the "reduced tensor" v associated to V 
\be
\label{rt}
v \equiv J(A)V
\ee
which has the  property of being invariant under the 
adapted diffeomorphism (\ref{ad}).
\be
\label{rtt}
v'= J(A-\d \Lambda)J(\d\Lambda)V = v
\ee

It makes sense to think about reduced tensors as the ones
in a D-dimensional manifold with a dimension locally shrunk
to a point. They keep their indices corresponding to the 
colapsed dimension, which are inert  to the adapted
diffeomorphismes, but  they are sensitive to other
transformations, as we will see in the T-duality case.

Looking at the explicit form of the 
matrix J, it is clear that the operation giving "reduced tensors" 
commutes with linear combinations, tensor products, 
contraction and 
permutation of indices.

Before following with this logic development let us see the most 
significant example. If we have a Riemmanian manifold with n 
commuting Killings 
\[G_{\mu\nu} = \left(\begin{array}{cc}
G_{ab} & A_{ai} \\
A_{bj} & \,\,\,\,\,\,\hat{G}_{ij} + A_{ic}A_{jd}G^{cd} \\
\end{array}
\right) \]
\be
\label{lg}
{\cal L}_{k_a}G_{\mu\nu}=0
\ee
where $G^{ab}$ is the inverse matrix of $G_{ab}$. It is well 
known that the 
desired "transverse" gauge fields are 
\be
\label{tgf}
A^{a}_{i}(x^j) = G^{ab}A_{bi}(x^j)
\ee
Using the convention of writing the usual tensors in capital 
letters and 
the reduced ones in small letters, the reduced metric takes the 
simple
form:
\[g_{\mu\nu}= \left(\begin{array}{cc}
G_{ab} & 0 \\
0 & \hat{G}_{ij}\\
\end{array}
\right)
\]

If we keep ourselves with the conditions (\ref{lie}) and 
(\ref{liec0}), 
we can repeat the arguments given above and conclude there 
exists the 
corresponding reduced covariant derivative
\footnote{the reduced covariant derivative is denoted by 
$\D$ too.The 
distinction is made looking over what kind of tensor acts.}:
\be
\label{rcd}
\D v \equiv J(A)\D V
\ee

Taking account 
the explicit expression for J in every tensor 
representation, we can read off the reduced 
connection $\gamma$ \footnote{details in the Apendix}:
\be
\label{rc} 
\gamma\ic=J(-A)^{\alpha}_{\mu} J(-A)^{\beta}_{\nu}
J(A)^{\rho}_{\delta}(\Gamma^{\delta}_{\alpha\beta} - 
\d_{\alpha}J(A)^{\delta}_{\beta})
\ee
In an arbitrary choice of the 
"transverse" gauge field $A^{a}_{i}$ the resulting reduced 
connection 
could not have any advantage,
but with a "natural" choice,ie., (\ref{tgf}) in a 
Riemmanian manifold, 
it is a very simplified version of the usual one. As the most 
significant
example I will write the reduced connection for the 
Levi-Civita connection for the metric in  (\ref{lg})
\bea
\label{rlc}
&\g^{c}_{ab} = 0 ;   &\g^{i}_{ab}= 
-{1\over2}\hat{\d}^{i}G_{ab} \nonumber\\
&\g^{b}_{ia}=\g^{b}_{ai}= 
{1\over2}G^{bc}\d_{i}G_{ca};\,\,\,\,\,\,\,\,\,\,\,\,\,
\,\,\,\,\,\,\,\,
&\g^{k}_{ij}=\hat{\Gamma}^{k}_{ij}
\nonumber\\
&\g^{i}_{ja}=\g^{i}_{aj}= {1\over2}G_{ab}\hat{F}(A)^{b\,i}_{j}; 
\,\,\,\,\,\,\,&\g_{ij}^{a}= -{1\over2}F(A)_{ij}^{a}
\eea

where hatted objects are the ones calculated with the 
quotient metric $\hat{G}_{ij}$ \footnote{$\hat{F}_{i}^{bj}=
\hat{G}^{jk}F_{ik}^{b}$ ;$\hat{\Gamma}_{ij}^{k}=
{1\over2}\hat{G}^{kl}(\d_{i}\hat{G}_{lj}+\d_{j}\hat{G}_{il}
- \d_{l}\hat{G}_{ij})$}, 
and $F(A)^{a}_{ij}\equiv \d_{i}A^{a}_{j} - \d_{j}A^{a}_{i}$ 
is the field 
strength of the gauge fields. The usual Levi-Civita connection 
for this 
case is written in the Apendix and the comparison shows the great 
advantage of using the reduced one. In the case  of an arbitrary 
covariant derivation, we can get the corresponding reduced 
connection 
adding the reduced tensor of the additional one to the 
Levi-Civita connection (\ref{rc}).
 
Despite its simplicity, even in the simplest case, 
the reduced connection
(\ref{rlc})
has a very rich structure 
because the presence of torsion at the same time that a 
non-Levi-Civita 
symmetric part, both restricted by the necessary 
covariantly constancy of
the reduced metric.

With the definition giving above of reduced covariant derivative 
the operation that gives the reduced tensors commutes with the 
basic 
operations of the tensor calculus: linear combination, tensor 
product, contraction, permutation of indices and covariant 
derivation. That feature together with its simplicity is the 
reason 
to call the whole setting  the "reduced geometry".

The generic presence of torsion is the responsible 
for a little subtlety in
calculating the reduced curvature. To see that, let us start 
with the
Riemann-Christoffel curvature. 
Due to the
commutation of reduced covariant differentiation
with the reduced mapping, we should write:
\be
\label{rcest}
[\D_{\mu},\D_{\nu}]W_{\rho}=J^{;\lambda\delta\pi}_{\mu\nu\rho
}(A)[\D_{\lambda},
\D_{\delta}]w_{\pi}
\ee
for an arbitrary one-form W belonging to $\Omega$, 
and its reduced version w.

Substituing commutators in both sides and taking account 
the presence of
torsion in the second one, we get :
\be
\label{rc2}
R(\Gamma_{L-C})^{\sigma}_{\mu\nu\rho }W_{\sigma} =
J^{;\lambda\delta\pi}_{\mu\nu\rho }
(A)(R(\gamma_{l-c})^{\eta}_{\lambda\delta\pi}w_{\eta}+
2T(\gamma_{l-c})
_{\lambda\delta}^{\eta}\D_{\eta}
w_{\pi})
\ee
I denote $\gamma_{l-c}$ the reduced Levi-Civita connection 
(\ref{rlc})
, $T(\gamma_{l-c})$ the associated torsion and $\Gamma_{L-C}$ 
the Levi-Civita connection. At first glance it 
seems to exist an obstruction to identify the reduced 
curvature by the presence 
of the torsion term.The little paradox solves realizing 
the only non
vanishing torsion
is $\gamma^{a}_{ij}$ (\ref{rlc}) and therefore that 
contribution does not
contain
derivatives of  w (because $\d_{a}w =0$) and can be 
added to the standart
curvature. The resulting reduced curvature is :
\be
\label{rc3}
(r_{l-c})^{\eta}_{\lambda\delta\pi}= R(\gamma_{l-c})
^{\eta}_{\lambda\delta\pi}-2T(\gamma_{l-c})^{a}_{\lambda\delta}
(\gamma_{l-c})_{a\pi}^{\eta}
\ee
In the general case we can write the connection as $\Gamma = 
\Gamma_{L-C} + H$. Using the formula

\bea
\label{rc333}
&R(\Gamma + Q)_{\mu\nu\sigma}^{\rho}=
R(\Gamma)_{\mu\nu\sigma}^{\rho}+
\nonumber\\
 &\D_{\mu}Q_{\nu\sigma}^{\rho}-
\D_{\nu}Q_{\mu\sigma}^{\rho}- Q_{\mu\sigma}^{\alpha}
Q_{\nu\alpha}^{\rho} + Q_{\nu\sigma}^{\alpha}
Q_{\mu\alpha}^{\rho}+ 2T(\Gamma)_{\mu\nu}^{\alpha}
Q_{\alpha\sigma}^{\rho}
\eea

being $\D_{\mu}$ the covariant derivative calculated from 
the generic $\Gamma$ connection and $T(\Gamma)$ the
associated torsion, and using the properties of the 
reduced map, we get for the reduced curvature :

\be
\label{rc4}
r^{\eta}_{\lambda\delta\pi}= 
R(\gamma_{l-c}+h)^{\eta}_{\lambda\delta\pi} 
-2T(\gamma_{l-c})^{a}_{\lambda\delta} 
(\gamma_{l-c}+ h)^{\eta}_{a\pi}
\ee

where $h_{\mu\nu}^{\rho}$ is the reduced tensor corresponding to 
$H_{\mu\nu}^{\rho}$.

In the present work we are interested  in the basic $U(1)$ 
duality.There, the reduced 
metric 
\[{g}_{\mu\nu}= \left( \begin{array}{cc}
\G & 0\\
0 & \hat{G}_{ij}\\
\end{array}
\right)
\]

has the additional advantage of having almost a trivial 
transformation 
under T-duality (\ref{BF});

\[\t{g}_{\mu\nu}= \left( \begin{array}{cc}
{1\over\G} & 0\\
0 & \hat{G}_{ij}\\
\end{array}
\right)
\]

It is not relevant here to make an exhaustive catalogue of the 
reduced geometry.The reduced exterior differentiation and the 
reduced 
Lie-derivative mapping can be obtained writing them in terms of 
covariant
derivatives(for example with the Levi-Civita connection);
in such a  way 
the reduced expression is manifest across (\ref{rcd}) and 
coincides with 
the usual one except for the presence of torsion in the reduced 
Levi-Civita
connection.The case of the differential of a two-form, 
which is relevant to the  $U(1)$ T-duality calculations is done.

Given a two-form B, its exterior differential, can be written in 
terms of the covariant one calculated with the Levi-Civita 
connection as

\be
\label{H}
H_{\mu\nu\rho}\equiv {1\over2}(\D_{\mu}B_{\nu\rho}+
\D_{\nu}B_{\rho\mu}+
\D_{\rho}B_{\mu\nu})
\ee

In those terms, the reduced H, i.e. h, is 
\bea
\label{h}
&h_{\mu\nu\rho}\equiv {1\over2}(\D_{\mu}b_{\nu\rho}+
\D_{\nu}b_{\rho\mu}+
\D_{\rho}b_{\mu\nu})=\nonumber\\
&{1\over2}(\d_{\mu}b_{\nu\rho}+\d_{\nu}b_{\rho\mu}+
\d_{\rho}b_{\mu\nu})- T(\gamma_{l-c})^{0}_{\mu\nu}b_{0\rho}-
T(\gamma_{l-c})^{0}_{\nu\rho}b_{0\mu}-
T(\gamma_{l-c})^{0}_{\rho\mu}b_{0\nu}
\eea
 
As we see the expression  of the reduced version 
in terms of ordinary 
derivatives gives an additional term due to the presence 
of torsion in 
the reduced connection. Despite this apparent setback 
the reduced H has
again nice properties from the T-duality point of view:
 
\bea
\label{h2}
&h_{0ij}= &-{1\over2}F(b)_{ij}\\
&h_{ijk}=&\hat{h}_{ijk}
\eea
where $F(b)_{ij}\equiv F( B_{0k}=b_{0k}= \t 
G_{0k}/\t \G = \t A_{k})_{ij}$\footnote{$\hat{h}_{ijk}=
{1\over2}(\d_{i}\hat{b}_{jk}+\d_{k}\hat{b}_{ij}+
\d_{j}\hat{b}_{ki})+{1\over4}(F(g)_{ij}B_{0k}+ 
F(b)_{ij}{G_{0i}\over\G}+
F(g)_{ki}B_{0j}+ F(b)_{ki}{G_{0j}\over\G}+
F(g)_{jk}B_{0i}+ F(b)_{jk}{G_{0i}\over\G})$
and $\hat{b}_{ij}\equiv B_{ij}-(G_{0i}B_{0j}-G_{0j}B_{0i})/2\G$
is the T-duality invariant $\tilde{\hat{b}}_{ij}=\hat{b}_{ij}$ 
transverse torsion potential.}. I
label the ijk component with a hat because it is T-duality 
selfdual:

\bea
\label{th}
&\t{h}_{0ij}=& -{1\over2}F(g)_{ij}\\
&\t{\hat{h}}_{ijk}=&\hat{h}_{ijk}
\eea

where $F(g)_{ij}=F(A_{k}=G_{0k}/\G)$.

The main purpose of the remainig sections is to exploit 
the power of the 
reduced framework in its application  for the study of 
the classical geometry
of the 
T-duality mapping.

\section{Generalized T-Duality Mapping}

The natural connections defined in the context of T-duality are 
$\Gamma^{\pm}$, with their reduced partners $\gamma^{\pm}$  

\bea
\label{gct}
&\Gamma^{\pm\rho}_{\mu\nu} =& \Gamma(L-C)\ic \pm 
H_{\mu\nu}^{\rho}\\
&\gamma^{\pm\rho}_{\mu\nu} =& \gamma(l-c)\ic \pm 
h_{\mu\nu}^{\rho}
\eea

where the torsion $ H_{\mu\nu\sigma}= 
H_{\mu\nu}^{\rho}G_{\rho\sigma}$ 
is as in (\ref{h2}).The explicit expressions for the 
reduced connection are gratefully simplified
\footnote{$\hat{h}_{ij}^{k}=
\hat{h}_{ijl}\hat{G}^{lk}$}:  

\bea
\label{rgc}
&\g^{\pm \,0}_{00} = 0 \nonumber\\
   &\g^{\pm \,i}_{00}= 
-{1\over2}\hat{\d}^{i}\G \nonumber\\
&\g^{\pm\, 0}_{i0}=\g^{\pm\,0}_{0i}= 
{1\over2}\d_{i}\ln \G\nonumber\\
&\g^{\pm\,k}_{ij}=\hat{\Gamma}^{k}_{ij}\pm
\hat{h}_{ij}^{k}
\nonumber\\
&\g^{\pm\,i}_{ 0j}= {1\over2}(\G\hat{F}(g)^{\,\,\,i}_{j}\mp
\hat{F}(b)^{\,\,\,i}_{j})
\nonumber\\
&\g^{\pm\,i}_{j0}= {1\over2}(\G\hat{F}(g)^{\,\,\,i}_{j}\pm
\hat{F}(b)^{\,\,\,i}_{j})
\nonumber\\
&\g_{ij}^{\pm\,0}= -{1\over2}(F(g)_{ij}\pm \G^{-1}{F}(b)_{ij})
\eea

Now it is easy to read off their T-duality 
transformation, that results to be diagonal:

\bea
\label{Trgc}
&\t\g^{\pm \,i}_{00}= -({1\over\G})^2\g^{\pm \,i}_{00} 
\nonumber\\
&\t\g^{\pm\, 0}_{i0}=\t\g^{\pm\,0}_{0i}= -\g^{\pm\, 0}_{i0}
\nonumber\\
&\t\g^{\pm\,i}_{ 0j}= \mp {1\over\G} \g^{\pm\,i}_{ 0j}
\nonumber\\
&\t\g^{\pm\,i}_{j0}= \pm{1\over\G}\g^{\pm\,i}_{j0}
\nonumber\\
&\t\g_{ij}^{\pm\,0}=  \pm\G \g_{ij}^{\pm\,0}\nonumber\\
&\t\g^{\pm\,k}_{ij}=\g^{\pm\,k}_{ij}
\eea

Except  for the $\gamma^{\pm \,0}_{i0}$ component, 
we can arrange 
it in the way

\be
\label{Trgc11}
\t \gamma^{\pm\rho}_{\mu\nu}= 
(-1)^{g_{\mu}}(\pm\G)^{(n^{0}-n_{0})}
\gamma^{\pm\rho}_{\mu\nu}
\ee

where I define $g_{0}\equiv1$ if the $\mu$ index 
is contravariant, 
$g_{0}\equiv-1$  if it is covariant and $g_{i}\equiv0$ 
in both cases;
Moreover $(n^{0}-n_{0})$
in front of 
an object (connection or tensor's component) 
with indices $\alpha$, means
$\sum_{ \alpha}^{ } g_{\alpha}$\footnote{I call it $n^0 - n_0 $
because it reduces to the number of contravariant 
components being zero 
minus the number of zero covariant ones.}.
The $\gamma^{\pm \,0}_{i0}$ component transforms flipping 
the sign 
with respect to (\ref{Trgc11}), 
which will be relevant in what follows.

It will be important too, to notice  
$\gamma^{\pm \,0}_{i0}$ acts as a
connection for (\ref{Trgc11})-type transformations : 
let the generic
T-duality 
transformation in the reduced setting $\t \Theta^{\pm} = 
(\pm\G)^{\Delta_{\Theta^{\pm}}}
\Theta^{\pm}$ for a given 
function $\Theta^{\pm}(X^j)$, 
there is a natural covariant derivative :

\be
\label{tdcd}
d_{ i}\Theta^{\pm}\equiv (\d_{i}+ 
\Delta_{\Theta^{\pm}}\gamma^{\pm\,0}_{i0})\Theta
^{\pm}
\ee

transforming as the $\Theta^{\pm}$ itself
\footnote{Moreover $d_{i}$
satisfies
the Leibniz's rule $d_{i}(A*B)=(d_{i}A)*B+A*(d_{i}B)$ 
and the "covariant
constancy" of  $\G$,i.e., $d_{i}\G=0$.}

\be
\label{tdcd2}
\t d_{ i}\t \Theta^{\pm} = 
(\pm\G)^{\Delta_{\Theta^{\pm}}}d_{i}\Theta^{\pm}
\ee

The simplicity of the reduced connection's 
transformation allow us to
realize there is a  diagonal T-duality transformation 
of reduced tensors
that maps (again diagonally in the reduced setting) the usual
(target-space)
$\D^{\pm}_{\mu}$ covariant derivatives. To see this, 
let us write them in
terms
of
the (\ref{tdcd}) :

\bea
\label{tdcd3}
\D_{0}^{\pm}v^{\ict}_{\icv}=
\{-\sum_{s=1}^{m}\g_{0\nu_{s}}^{\pm\sigma}
v^{\ict}_{\nu_1 ,..,\sigma,..,\nu_{m }} +
\sum_{r=1}^{l}\g_{0\sigma}^{\pm\mu_{r}}
v_{\icv}^{\mu_1 ,..,\sigma,..,\mu_{l }}\} \nonumber\\
\D_{i}^{\pm}v^{\ict}_{\icv}= \{d_{i}v^{\ict}_{\icv}
-\sum_{s=1/\nu_{s}\neq0}^{m}\g_{i\nu_{s}}^{\pm\sigma}
v^{\ict}_{\nu_1 ,..,\sigma,..,\nu_{m }} +
\sum_{r=1/\mu_{r}\neq0}^{l}\g_{i\sigma}^{\pm\mu_{r}}
v_{\icv}^{\mu_1 ,..,\sigma,..,\mu_{l }}-\nonumber\\
\sum_{s=1/\nu_{s}=0}^{m}\g_{i0}^{\pm\,j}
v^{\ict}_{\nu_1 ,..,j,..,\nu_{m }} + \sum_{r=1/\mu_{r}=0}^{l}
\g_{ij}^{\pm\,0}
v_{\icv}^{\mu_1 ,..,j,..,\mu_{l}}\}+\nonumber\\
((n^{0}-n_{0}) - \Delta_{v_{\icv}^{\ict}})
\gamma^{\pm \,0}_{i0} v^{\ict}_{\icv}
\eea

The last term in (\ref{tdcd3}) transforms with an 
undesired $-1$ with respect to the dominant  $d_{i}v$, 
fixing the
weight $\Delta_{v_{\icv}^{\ict}}= (n^{0}-n_{0})$.
Therefore the reduced transformation is 

\be
\label{tcd}
\t v^{\pm\ict}_{\icv} = (\pm\G)^{(n^{0} -n_{0})}v^{\ict}_{\icv}
\ee

giving the linear map for the covariant derivatives 
under T-duality
\footnote{It must be stressed, to do not overcarry the notation,
I always write $n^{0}-n_{0}$, but in 
every case its value is given by the tensor's 
(connection's) component  in
front of it 
in the way  described above.}

\be
\label{tcd2}
\t\D_{\rho}^{\pm}\t v^{\pm\ict}_{\icv}= (-1)^{g_{\rho}} 
(\pm\G)^{(n^{0} -n_{0})^{'}}\D_{\rho}^{\pm}v^{\ict}_{\icv}
\ee

Before writing the transformations in the 
usual (non-reduced)setting, I express them 
in a compressed notation as 

\bea
\label{30}
&\t v^{\pm} \equiv &D_{\pm}(\G) v \\
&\t\D^{\pm}\t v^{\pm} \equiv &D_{\D}^{\pm}(\G) \D^{\pm} v
\eea

$D_{\pm}$ and $D^{\pm}_{\D}$ being diagonal matrices 
in every tensor 
representation.

Inverting (\ref{rt}) for $v$ and $\t v^{\pm}$ 
taking account $A_i = 
G_{0i}/\G$ and $\t A_i = B_{0i}$ we get for 
$\t V^{\pm}$ and $\t \D^{\pm}
\t V^{\pm}$ : 
\bea
\label{31}
&\t V^{\pm} = & J(-B_{0i})D_{\pm}(\G)J(G_{0i}/\G)V \\
& \t\D^{\pm} \t V^{\pm} = & 
J(-B_{0i})D_{\D}^{\pm}(\G)J(G_{0i}/\G)\D^{\pm} V
\eea
Because J, $D_{\pm}$ and $D^{\pm}_{\D}$ factorise,
the T-duality mapping does too and can be 
written in terms of the matrices $T^{\pm}$ and $T_{\pm}$ 
defined below as:

\be
\label{32}
\t V^{\pm\ict}_{\icv} = 
(\prod_{r=1}^{l} T^{\mu_r}_{\pm\,\beta_r}) 
(\prod_{s=1}^{m} T^{\pm\,\alpha_s}_{\nu_s})
V^{\beta_1 ,...,\beta_l }_{\alpha_1 ,...,\alpha_m }
\ee

\be
\label{32222222}
\t\D_{\rho}^{\pm}\t V^{\pm\ict}_{\icv} = 
T^{\mp\,\lambda}_{\rho}(\prod_{r=1}^{l}T^{\mu_r}_{\pm\,\beta_r})
(\prod_{s=1}^{m}T^{\pm\,\alpha_s}_{\nu_s})
\D_{\lambda}^{\pm}V^{\beta_1 ,...,\beta_l }_{\alpha_1 
,...,\alpha_m }
 \ee

with

\[ T^{\pm\,\nu}_{\mu} = \left ( \begin{array}{cc}
\pm{1\over\G} & 0 \\
(\pm B_{0i} - G_{0i})/\G\,\,\,\, & \delta^{i}_{j}
\end{array}
\right ) \]

\[ T_{\pm\,\nu}^{\mu} = \left ( \begin{array}{cc}
\pm \G\,\,\,\, &\pm G_{0i} - B_{0i}  \\
0 & \delta^{i}_{j}
\end{array}
\right ) \]

where $\nu$ is the column index and $\mu$ is the row index.
These matrices $T_{\pm}$ and  $T^{\pm}$,
introduced by Hassan  to the study of
the T-duality of the extended supersymmetry \cite{H}, 
can be thought as a sort
of "vielvein" relating indices 
of the initial and dual geometries.
In what follows I will call (\ref{32}) T-duality canonical 
transformation for the tensor V. Let us note (\ref{32222222})
is not canonical. This anomaly is the 
responsible for the generalized curvatures of original 
and dual geometries 
do not transform simply changing indices with the 
$T^{\pm}$ vielveins. 
We will see it in detail in the last section.

It must be stressed there are other tensors whose T-duality 
transformation is not as in (\ref{32});
the most appealing cases are the 
torsion potential $B_{\mu\nu}$ and its field strength, 
the torsion 
$H_{\alpha\beta\gamma}$ \cite{H}.
It seems the natural place for the torsion is 
taking part of the generalized connection; then from 
(\ref{32})(\ref{32222222}) or (\ref{Trgc}) 
we can extract the whole 
transformation to be :

\be
\label{34}
\t \Gamma^{\pm\,\rho}_{\mu\nu} = T^{\mp\,\lambda}_{\mu}
T^{\pm\,\beta}_{\nu}
T_{\pm\,\alpha}^{\rho} \Gamma^{\pm\,\alpha}_{\lambda\beta} + 
(\d_{\mu} T^{\pm\,\beta}_{\nu}) T_{\pm\,\beta}^{\rho}
\ee

For tensors covariantly constant with respect to one 
of both derivatives
$\D^{\pm}_{\mu}$, the T-duality mapping is enforced to be 
given by (\ref{32}).
This is the case of the metric itself because 
$\D^{\pm}_{\mu} G = \t \D^{\pm}_{\mu} 
\t G = 0 $. If we have in the manifold two covariantly 
constant p-forms 
$A^{\pm}$ satisfying $\D^{\pm}_{\mu} A^{\pm} = 0$,  
there is a W-algebra in the underlying string 
sigma-model.\footnote{That includes the mapping 
of complex 
structures , although it requires the additional vanishing of 
the Nijenhuis tensor \cite{H}}Another dual W-algebra is present 
in the dual 
string theory provided the p-forms transforms as in 
(\ref{32})\cite{K}:

\bea 
\label{33} &\t A^{\pm}_{0 i_1 ... i_{p-1}} = &\pm{1\over\G}
A_{0 i_1  ...
i_{p-1}}^{\pm}\nonumber\\
&\t A^{\pm}_{i_1 ... i_p } = & A^{\pm}_{i_1 ... i_p } + 
\sum^{p}_{s=1}{(\pm
b_{0i_{s}} - g_{0i_{s}})\over\G}A_{i_1 ..0..i_p }
\eea

If we have another Killing $K^{(2)}_{\mu}$ 
(non necessarily commuting
with the one used to T-dualize) 
giving rise to a chiral current, it holds 
\cite{RV} $\D_{\mu}^{\pm}K^{(2)}_{\nu}=0$ 
($+$ for an holomorphic 
current and $-$ for an antiholomorphic one). The T-duality 
canonical transformation  of $K^{(2)}_{\mu}$ gives
another dual Killing vector with chiral current.

Instead of the privileged T-duality mapping for the 
generalized covariant
derivative,
we could be interested in the one for the "generalized 
divergence" 
$\D_{\lambda}^{\pm}Q^{\lambda\,\ict}_{\pm\,\icv}$. Again the 
study is very simplified in the reduced framework. 

After determining the T-duality weight of 
$q^{\lambda\ict}_{\pm\,\icv}$
following 
the same procedure as in the covariant derivatives' map, 
the required mapping results to be :

\bea
\label{3333}
&\t q^{\lambda\,\ict}_{\pm\,\icv} = (-1)^{g_{\lambda}}
(\pm\G)^{(n^{0} -n_{0}+1)}q^{\lambda\,\ict}_{\icv}\nonumber\\
&\t \D_{\rho}^{\pm}\t q^{\rho\,\ict}_{\pm\,\icv}= 
(\pm\G)^{(n^{0}
-n_{0})^{'}+1}\D_{\lambda}^{\pm}q^{\lambda\,\ict}_{\icv}
\eea

We have learned to read it in the common language as
($K^2 = K_{\mu}K^{\mu}$)
\bea
\label{333}
&\t Q^{\lambda\,\ict}_{\pm\,\icv}=
K^2T^{\lambda}_{\mp\,\rho}(\prod_{r=1}^{l}T^{\mu_r}_{\pm\,
\beta_r})
(\prod_{s=1}^{m}T^{\pm\,\alpha_s}_{\nu_s})
Q^{\rho\,\beta_1 ,...,\beta_l }_{\alpha_1 ,...,\alpha_m }
\nonumber\\
&\t \D_{\lambda}^{\pm}\t Q^{\lambda\,\ict}_{\pm\,\icv} = 
K^2(\prod_{r=1}^{l} T^{\mu_r}_{\pm\,\beta_r}) 
(\prod_{s=1}^{m} T^{\pm\,\alpha_s}_{\nu_s})
\D_{\rho}^{\pm}Q^{\rho\,\beta_1 ,...,\beta_l }_{\alpha_1 
,...,\alpha_m } 
\eea

This section shows how elementary geometrical 
objects transforms under T-duality
depending on their defining properties
: covariantly constant tensors
transforms canonically 
(\ref{32}),i.e., changing indices with the "vielbeins" 
$T_{\pm}$ and $T^{\pm}$ ;
"divergenceless" tensors do it under 
(\ref{333}) changing the index
corresponding with the diveregence with 
$T_{\mp}$ instead of $T_{\pm}$ ; 
the generalized connection
transforms as a true connection (\ref{34})
with respect to the T-duality
canonical transformation with the only peculiarity of
using 
$T_{\mp}$ instead of $T_{\pm}$ in the index corresponding to the
derivation. This anomaly 
propagates to the T-duality transformation 
of every index associated to
derivation as we have seen
in (\ref{32})(\ref{333}), and finally
it will be the responsible for the
inhomogeneus transformation of the generalized curvature, 
as we will show in
the fifth section.


\section{T-Duality Classical Dynamics}

In addition to the D-dimensional manifold M representing the
target space-time, in the context of strings 
we have a two dimensional
world embedded in it, 
the world-sheet $\Sigma$, identified with the dinamical string
\footnote{ I omit here p-branes, D-branes and any other kind 
of stringy
extended objects.}. At 
tree-level in the string dinamics $\Sigma$ 
has the topology of the sphere.
Choosing light-cone
real coordinates $\sigma^{\pm}$, 
the covariant world-sheet derivatives for
mappings $Y(\sigma^{+},\sigma^{-})$between
$\Sigma$ and the tangent space of our manifold M in
$X^{\mu}(\sigma^{+},\sigma^{-})$are:

\be
\label{40}
\D_{\pm} = \d_{\pm} + \Gamma_{\pm}
\ee

being  the pull-back in terms of the string's embedding 
$X^{\mu}(\sigma^{+},\sigma^{-})$ 

\be
\label{41}
\Gamma_{\pm\, \mu}^{\rho} =
\d_{\pm}X^{\alpha}\Gamma^{\mp\,\rho}_{\alpha\mu}
\ee

Extending the definition (\ref{rt}) to the $Y$ mappings 

\be
\label{422}
y(\sigma^{+},\sigma^{-})\equiv J(A(X(\sigma^{+},\sigma^{-})))
Y(\sigma^{+},\sigma^{-})
\ee

where $y$ is the reduced mapping. 
With that definition we can proof the
reduced partner
of the pull-back is 

\be
\label{rpb}
\g^{\rho}_{\pm\mu}= (\d_{\pm}x)^{\beta}\g^{\mp\rho}_{\beta\mu}
\ee

where I call $ (\d_{\pm}x)^{\beta}\equiv
J(A)^{\beta}_{\nu}\d_{\pm}X^{\nu}$ following
(\ref{422})\footnote{I write the reduced $\d X^{\mu}$ 
between parentheses to
specify
that in general it is not the partial derivative of anything. 
Only when the pull-back of 
$F(A)_{ij}$ vanishes $x^{\mu}$ can be 
identified with a sort of  U(1)
invariant embedding coordinates.}.

In the reduced framework it is easy to convince ourselves 
that the only
T-dual
change for  $(\d_{\pm}x)$ that transforms the pull-back 
diagonally is
nothing just
the one responsible for the Buscher 's formulas : 

\be
\label{42}
\d_{\pm}\t X^{\mu} = T^{\mu}_{\pm\,\,\nu}\d_{\pm}X^{\nu}
\ee

being the reduced pullback transformation 

\bea
\label{43}
&\t \gamma_{\pm\,\,0}^{0}= 
&- \gamma_{\pm\,\,0}^{0}\nonumber\\
&\t \gamma_{\pm\,\,i}^{0}= 
&\mp\G \gamma_{\pm\,\,i}^{0}\nonumber\\
&\t \gamma_{\pm\,\,0}^{i}= 
&\mp{1\over\G} \gamma_{\pm\,\,0}^{i}\nonumber\\
&\t \gamma_{\pm\,\,j}^{i}= 
&{1\over\G} \gamma_{\pm\,\,j}^{i}
\eea

which again can be summarized 

\be
\label{abcde}
\t \gamma_{\pm\,\,\mu}^{\nu}= 
(\mp\G)^{(n^{0}-n_{0})} 
\gamma_{\pm\,\,\mu}^{\nu}
\ee

except for the $\gamma_{\pm\,\,0}^{0}$, 
in which there is a flip of sign
allowing the T-duality covariantization (\ref{tdcd}) 
of the world-sheet 
covariant derivatives.

It is worthwhile to mention (\ref{42}) implies 
$ K_{\mu}\D_{+}\D_{-}
X^{\mu}=
K_{\mu}\D_{-}\D_{+} X^{\mu}=0
\longleftrightarrow \t K_{\mu}\t \D_{+} \t \D_{-} \t X^{\mu}= 
\t K_{\mu}\t
\D_{-} \t \D_{+}
\t X^{\mu}=0$, which is automatically valid 
for the classical string 
(it is the current conservation corresponding to the isometry).

Again the simplicity of (\ref{43}) allow us to 
build the diagonal mapping for any $Y(\sigma^{+},\sigma{-})$ 
mapping 
\footnote{The intermediate tool is now the analogue to 
(\ref{tdcd}) 
in the world-sheet,i.e., 
$d_{\pm}\Theta(\sigma^{+},\sigma^{-})\equiv
 (\d_{\pm}+ \Delta_{\Theta}\gamma_{\pm \,0}^{0})\Theta
(\sigma^{+},\sigma^{-})$, transforming as $\Theta$ with weight 
$\Delta_{\Theta}$ under T-duality.} :

\be
\label{44}
\t y^{\mp}(\sigma^{+},\sigma^{-})^{\ict}_{\icv} = 
(\mp\G)^{(n^{0}
-n_{0})}
y(\sigma^{+},\sigma^{-})^{\ict}_{\icv}
\ee

with the property of transforming linearly the covariant 
derivatives
$\D_{\pm}$:

\be
\label{45}
\t \D_{\pm}\t y(\sigma^{+},\sigma^{-})^{\mp\ict}_{\icv}=
(\mp\G)^{(n^{0}-n_{0})}
\D_{\pm} y(\sigma^{+},\sigma^{-})^{\ict}_{\icv}
\ee

In the usual setting (\ref{44}) and (\ref{45}) 
can be written with the help
of the 
preceding work (\ref{tcd}) (\ref{tcd2}) and (\ref{32}) 
(\ref{32222222}) as

\bea
\label{46}
&\t Y(\sigma^{+},\sigma^{-})^{\pm\ict}_{\icv} = 
&(\prod_{r=1}^{l} T^{\mu_r}_{\pm\,\beta_r}) 
(\prod_{s=1}^{m} T^{\pm\,\alpha_s}_{\nu_s})
Y(\sigma^{+},
\sigma^{-})^{\beta_1 ,...,\beta_l }_{\alpha_1 ,...,\alpha_m
}\nonumber\\
&\t\D_{\mp}\t Y^{\pm\ict}_{\icv} =& 
(\prod_{r=1}^{l}T^{\mu_r}_{\pm\,\beta_r})
(\prod_{s=1}^{m}T^{\pm\,\alpha_s}_{\nu_s})
\D_{\mp}Y^{\beta_1 ,...,\beta_l }_{\alpha_1 ,...,\alpha_m }
\eea

(\ref{46}) allow us to extract the pull-back's 
T-duality transformation in the
common language:

\be
\label{47}
\t \Gamma^{\rho}_{\pm\mu} = T^{\mp\,\lambda}_{\mu}
T^{\rho}_{\mp\,\beta} \Gamma^{\beta}_{\pm\lambda} + 
(\d_{\pm} T^{\mp\,\beta}_{\mu}) T_{\mp\,\beta}^{\rho}
\ee

The most relevant example of this mapping is provided by the
$\d_{\pm}X^{\mu}$
for which holds 
$\t \D_{\pm}\d_{\mp}\t X^{\mu}=  
T_{\mp\nu}^{\mu}\D_{\pm}\d_{\mp}X^{\nu}$,
giving  the classical stringy equivalence 
between the two different
geometries.

\section{Generalized Curvature's Transformation}
Defining as  usual the generalized curvature as 
$R^{\pm}_{\mu\nu\sigma\rho}
= R(\Gamma^{\pm})^{\lambda}_{\mu\nu\sigma} G_{\lambda\rho}$ 
which
has the symmetry properties $ R^{\pm}_{\mu\nu\sigma\rho}= -
R^{\pm}_{\nu\mu\sigma\rho}
= - R^{\pm}_{\mu\nu\rho\sigma}$, $ R^{\pm}_{\mu\nu\rho\sigma}=
R^{\mp}_{\rho\sigma \mu\nu}$ and taking account the generalized
connection's 
transformation (\ref{Trgc})
\footnote{See Apendix for explicit formulas}
we get the dual generalized curvature in the
reduced framework :

\bea
\label{50}
& \t r^{\pm}_{0i0j} =& -{1\over(\G)^2}( r^{\pm}_{0i0j} -
{1\over2\G}\d_{i}\G
\d_{j}\G)
 \nonumber\\
& \t r^{\pm}_{0ijk} =& \mp{1\over\G}( r^{\pm}_{0ijk} - \d_{i}\G
\gamma^{\mp\,
0}_{jk})
 \nonumber\\
& \t r^{\pm}_{ijk0} =& \pm{1\over\G}( r^{\pm}_{ijk0} + \d_{k}\G
\gamma^{\pm\,
0}_{ij})
 \nonumber\\
& \t r^{\pm}_{ijkl} =& r^{\pm}_{ijkl} - 2\G \gamma^{\pm\, 0}_{ij}
\gamma^{\mp\, 0}_{kl}
\eea

We can convince ourselves the inhomogeneus 
part of the transformation
can be writen in terms of the Killing vector, 
giving the compact result :

\be
\label{501}
\t r^{\pm}_{\mu\nu\sigma\rho}= 
(-1)^{(g_{\mu} + g_{\nu})}(\pm\G)^{-n_{0}}
(r^{\pm}_{\mu\nu\sigma\rho}-{2\over k^2}\D_{\mu}^{\pm}k_{\nu}
\D_{\sigma}^{\mp}k_{\rho})
\ee

where $K^2=k^2=K_{\mu}K^{\mu}=\G$. 
In the usual setting the transformation (\ref{501}) reads
\footnote{$\D_{\mu}^{\pm}k_{\nu}$ only has antisymmetric part
due to the Killing condition.}

\be
\label{5031}
\t R^{\pm}_{\mu\nu\sigma\rho} = 
T^{\mp\,\alpha}_{\mu} T^{\mp\,\beta}_{\nu}
T^{\pm\,\delta}_{\sigma} T^{\pm\,\eta}_{\rho}(
R^{\pm}_{\alpha\beta\delta\eta}
-{2\over K^2}\D_{\alpha}^{\pm}K_{\beta}
\D_{\delta}^{\mp}K_{\eta} )
\ee

It is important to note that the inhomogeneus  
part of the transformation
only depends on the 
transverse components $G_{0\,\mu}, B_{0i}$ and their first
skew-symmetric derivatives.

When an object  transforms inhomogeneusly, say 
$\t r = \pm (\G)^{\Delta}( r + \psi )$
, the involution property of the T-duality transformation  
$T^2=1$
fixes the $\psi$ transformation to be 
$\t \psi = \mp (\G)^{\Delta}\psi$,
allowing  to create the homogeneus 
$ w \equiv r + {1\over2} \psi $
i.e.,  $\t w =  \pm (\G)^{\Delta} w $. Specifically, it means

\be
\label{qwrtdesx}
{1\over \t K^2}\t \D_{\mu}^{\pm}\t K_{\nu}
\t \D_{\sigma}^{\mp}\t K_{\rho} 
 = - \,T^{\mp\,\alpha}_{\mu} T^{\mp\,\beta}_{\nu}
T^{\pm\,\delta}_{\sigma} T^{\pm\,\eta}_{\rho}(
{1\over K^2}\D_{\alpha}^{\pm}K_{\beta}
\D_{\delta}^{\mp}K_{\eta} )
\ee

and therefore we can create 
the "corrected " generalized curvature
$W_{\mu\nu\sigma\rho}^{\pm}$ 
transforming linearly under T-duality :

\bea
\label{530}
&W^{\pm}_{\mu\nu\sigma\rho}\equiv R^{\pm}_{\mu\nu\sigma\rho}
-{1\over K^2}\D_{\mu}^{\pm}K_{\nu}
\D_{\sigma}^{\mp}K_{\rho}\nonumber\\
& \t W^{\pm}_{\mu\nu\sigma\rho}= T^{\mp\,\alpha}_{\mu} 
T^{\mp\,\beta}_{\nu}
T^{\pm\,\delta}_{\sigma} T^{\pm\,\eta}_{\rho}
W^{\pm}_{\alpha\beta\delta\eta}
\eea

At this point is for free the rederivation of  the
generalized Ricci-tensor's transformation giving the need of 
a non-trivial dilaton change under T-duality in order 
to guarantee 
the one-loop conformal invariance of the dual string sigma-model 
\cite{BU}.

We get  the dual-reduced-Ricci-tensor from (\ref{50})
$r_{\mu\nu}^{\pm}$:

\be
\label{5000x}
\t r_{\mu\nu}^{\pm}= (-1)^{g_{\mu}}(\pm\G)^{-n_{0}}
( r_{\mu\nu}^{\pm}- \D_{\mu}^{\pm}\D_{\nu}^{\pm}
\ln \G)
\ee

It is well known the sigma-model coupling between the two-
dimensional  curvature and an scalar field called dilaton,
$\Phi$,  it is enough 
to ensure the vanishing of the dual one-loop beta function, 
provided the former transforms under T-duality as 

\be
\label{5000xx}
\t \Phi = \Phi - {1\over2}\ln\G
\ee

In other words, the tensor representing 
the one-loop beta function
for the string sigma-model (bosonic and supersymmetric) 
\cite{CLNY},ie,
$\beta_{\mu\nu}^{\pm}= R_{\mu\nu}^{\pm}- 2\D_{\mu}^{\pm}
\D_{\nu}^{\pm}\Phi$, transforms linearly under T-duality
as can be read off from (\ref{5000x}) and (\ref{5000xx}) :

\be
\label{500xxx}
\t \beta^{\pm}_{\mu\nu}= T^{\mp\,\alpha}_{\mu} 
T^{\pm\,\lambda}_{\nu}
\beta^{\pm}_{\alpha\lambda}
\ee

Therefore the vanishing of 
$\beta_{\mu\nu}$ implies the one for 
$\t \beta_{\mu\nu}$ and vice versa.
Following this approach, the dilaton one-loop beta 
function can be obtained
trying to complete minimally the generalized scalar curvature 
$R^{\pm}$ 
in order to get a T-duality scalar ; the transformation for
$R^{\pm}=r^{\pm}$
results to be (\ref{50}) (\ref{5000x}) :

\be
\label{5xyz}
\t R^{\pm}=\t r^{\pm}= R^{\pm} -2\hat{\D}^{\pm\,i}\d_{i}\ln\G
+2{1\over\G}\g_{i0}^{\pm\,k}\g_{0k}^{\pm\,i}
\ee

and the desired T-duality scalar is

\be
\label{5xyzw}
\beta^{\Phi}= R^{\pm}+4((\d \Phi)^2 - (\D^{\pm})^2 \Phi) 
- {2\over3}
H^2
\ee

where we define $H^2_{\alpha\beta}\equiv H_{\alpha\nu\rho} 
H^{\,\,\,\,\nu\rho}_{\beta}$. Modulo a 
constant term it coincides with the one-loop dilaton 
beta-function
\cite{CLNY}.
\section{Canonical Connection and Curvature}
The anomalous transformation of the index corresponding with 
first covariant derivations propagates in an annoying way to 
objects constructed from higher derivations, such as the 
generalized curvatures and the one-loop beta functions. 
In other words, the covariant derivation $\D^{\pm}$ does not 
commute with the canonical map defined in (\ref{32}). 

Looking at the generalized connections' transformation 
(\ref{Trgc}) we 
notice there is a minimal covariant subtraction 
giving a connection,
the canonical connection, which commutes with 
the canonical T-duality:

\bea
\label{tdcc1}
&\bar{\Gamma}_{\mu\nu}^{\pm\,\,\rho}\equiv \Gamma_{\mu\nu}^
{\pm\,\,\rho}-{1\over K^2}K_{\mu}\D_{\nu}^{\mp}K^{\rho}
\nonumber\\
&\t{\bar{\Gamma}}^{\pm\,\rho}_{\mu\nu} = T^{\pm\,\lambda}_{\mu}
T^{\pm\,\beta}_{\nu}
T_{\pm\,\alpha}^{\rho} 
\bar{\Gamma}^{\pm\,\alpha}_{\lambda\beta} + 
(\d_{\mu} T^{\pm\,\beta}_{\nu}) T_{\pm\,\beta}^{\rho}
\eea

Because the anomaly mentioned above is located in the $\g_{0\nu}
^{\pm\,\,\sigma}$ components, the whole effect of the 
subtraction
is to cancel against them, giving 
$\bar{\g}_{0\mu}^{\pm\,\,\rho}=0$
and $\bar{\g}_{i\nu}^{\pm\,\,\rho}=\g_{i\nu}^{\pm\,\,\rho}
$\footnote{
These conditions guarantee that if a tensor is covariantly 
constant with
respect to $\D_{\mu}^{\pm}$ it does too with respect to 
$\bar{\D}_{\mu}^{\pm}$.The converse is in general not true.
Therefore the metric commutes with $\bar{\D}_{\mu}^{\pm}$}.This 
fact allows  the commutation between 
$\bar{\D}^{\pm}_{\mu}$ and the
canonical T-duality map :

\bea
\label{tdcc12x}
&\t V^{\pm\ict}_{\icv} = 
&(\prod_{r=1}^{l} T^{\mu_r}_{\pm\,\beta_r} 
\prod_{s=1}^{m} T^{\pm\,\alpha_s}_{\nu_s})
V^{\beta_1 ,...,\beta_l }_{\alpha_1 ,...,\alpha_m }\nonumber\\
&\t{\bar{\D}}_{\rho}^{\pm}\t V^{\pm\ict}_{\icv} =& 
T^{\pm\,\lambda}_{\rho}(\prod_{r=1}^{l}T^{\mu_r}_{\pm\,\beta_r})
(\prod_{s=1}^{m}T^{\pm\,\alpha_s}_{\nu_s})
\bar{\D}_{\lambda}^{\pm}V^{\beta_1 ,...,\beta_l }_{\alpha_1 
,...,\alpha_m }
 \eea

Another consecuence is 
$\bar{\D}_{0}^{\pm}=0$ implying 
$\bar{R}_{0\nu\sigma\rho}^{\pm}=0$.
Even more, the new connection is compatible with the metric
provided that $K^{\mu}$ is a Killing vector.

In a certain sense the barred connection 
seems to be the most natural
associated to the presence of a Killing. 
If we think the Killing as 
a vector field indicating the direction in which nothing 
changes, we would expect the parallel 
transport is really insensitive
to displacements in that direction. 
This  happens with the canonical connection but no
with the Levi-Civita (or its 
torsionfull generalizations) one.

The commutation with the canonical T-duality 
implies that the curvature 
for $\bar{\D}^{\pm}$ transforms canonically 
(the same as the Ricci tensor 
and the scalar curvature).

\be
\label{tdcc12}
 \t{\bar{R}}^{\pm}_{\mu\nu\sigma\rho}= T^{\pm\,\alpha}_{\mu} 
T^{\pm\,\beta}_{\nu}
T^{\pm\,\delta}_{\sigma} T^{\pm\,\eta}_{\rho}
\bar{R}^{\pm}_{\alpha\beta\delta\eta}
\ee
 I will list five independent T-duality scalars

\bea
\label{tcdc123}
&I_{1}= R^{\pm}-(\D^{\pm})^{2}\ln K^2- {2\over K^{2}}
H^{2}_{\alpha\beta}K^{\alpha}K^{\beta}\nonumber\\
&I_{2}= H^{2}-{3\over K^{2}}H^{2}_{\alpha\beta}
K^{\alpha}K^{\beta}\nonumber\\
&I_{3}= {1\over K^{2}}( \D_{\mu}K_{\nu}\D^{\mu}K^{\nu}
+ H^{2}_{\alpha\beta}K^{\alpha}K^{\beta})\nonumber\\
&I_{4}={1\over K^2}K_{\alpha}\d_{\beta}K_{\lambda} 
H^{\alpha\beta\lambda}\nonumber\\
&I_{5}=({\d K^{2} \over K^{2}})^{2}
\eea

built with the help of $\bar{R}^{\pm}$ and being $\D_{\mu}$ the
Levi-Civita covariant derivation. In particular, $\bar{R}^{\pm}
= I_{1}\mp 2I_{4}$.

Finally, with this covariant derivation, 
the T-duality canonical map
commutes with the basic geometrical operations: 
linear combinations, 
tensor products, permutation of indices, 
contractions and covariant
derivations.
\section{Conclusions}
This work shows how the reduced geometry is a 
privileged framework
to the study of the T-duality's geometry and 
possibly of many other
different issues related with the abelian Killing vectors.

The T-duality transformation diagonalizes in the reduced 
setting, 
allowing us to get in a straightforward way results pursued
since a long time,
such as the generalized curvatures' map, the canonical map for 
the covariant derivatives in the target-space and 
in the world-sheet,
the minimal correction to connections and curvatures 
in order to 
transform linearly and T-duality scalars.
The introduction of the dilaton can be seen
as the minimal 
modification needed to map the 1-loop beta-functions 
preserving conformal
invariance, but from the geometrical view, the dilaton is 
completely
insufficient to
build in a systematic way T-duality tensors
(i.e., tensors transforming canonically under
T-duality).The object serving to "covariantize"
under T-duality is the Killing vector itself across 
the canonical
connection,
as it was shown in the 
last section. In connection to that, new T-duality
scalars have been found without the help of the 
dilaton.
 
Future work could be the higher-loop corrections to the 
Buscher's formulas , the map for the invariants characterizing
the geometry and the topology of the manifold, the global
questions in the reduced setting,  the non-abelian
generalizations of this procedure, and the deeper
study of the geometry of the canonical connection and its
relation with the string sigma model.

\subsection*{Acknowledgements}
I wish to thank E. Alvarez  for suggesting, reading and 
commenting
this work. Also I want to thank P. Meessen for useful 
discussions.
This work has been partially supported by a C.A.M. grant.
\appendix
\section{Apendix :
The reduced  connection and Riemaniann curvature}
In order to have the opportunity of getting an idea about the
operativity of the reduced framework, I show here 
the usual Levi-Civita
connection for the  generic metric with n-commuting 
Killing vector fields :

\[G_{\mu\nu} = \left(\begin{array}{cc}
G_{ab} & A_{ai} \\
A_{bj} & \hat{G}_{ij} + A_{ic}A_{jd}G^{cd} \\
\end{array}
\right) \]

\bea
\label{1000}
&\Gamma^{a}_{bc}= {1\over2}A_{i}^{a}\hat{\d}^{i}G_{bc}
\nonumber\\
&\Gamma^{i}_{ab}= -{1\over2}\hat{\d}^{i}G_{ab}\nonumber\\
&\Gamma^{j}_{ai}=\Gamma^{j}_{ia}=
{1\over2}(G_{ab}\hat{F}_{i}^{b\,j}-A_{i}^{b}\hat{\d}^{j}
G_{ab})\nonumber\\
&\Gamma^{b}_{ai}=\Gamma^{b}_{ia}=
{1\over2}(G^{bc}\d_{i}G_{ac}-A_{j}^{b}G_{ac}
\hat{F}_{i}^{c\,j}+A_{i}^{c}A_{j
}^{b}\hat{\d}^{j}G_{ac})\nonumber\\
&\Gamma^{k}_{ij}=\hat{\Gamma}^{k}_{ij}+{1\over2}G_{ab}(
A_{j}^{b}\hat{F}_{i}^{a\,k}+A_{i}^{b}\hat{F}_{j}^{a\,k})
-{1\over2}A_{i}^{a}A_{j}^{b}\hat{\d}^{k}G_{ab}\nonumber\\
&\Gamma^{a}_{ij}={1\over2}\{(\hat{\D}_{i}A_{j}^{a}+
\hat{\D}_{j}A_{i}^{a})
+A_{i}^{c}A_{j}^{b}A_{k}^{a}\hat{\d}^{k}G_{bc}
+G^{ab}(A_{j}^{c}\d_{i}G_{bc}+ A_{i}^{c}\d_{j}G_{bc})-
A_{k}^{a}G_{cb}(A_{j}^{b}\hat{F}_{i}^{c\,k}
+A_{i}^{b}\hat{F}_{j}^{c
\,k})\}\nonumber\\
&  
\eea

The (\ref{1000}) expressions must be compared with 
(\ref{rlc}) to realize the
advantages.

Now I would like to make explicit the relation 
between a generic connection
 $\Gamma$ 
and its reduced version $\gamma$ :
\bea
\label{1100}
&\g^{i}_{ab}= \Gamma_{ab}^{i}\nonumber\\
&\g^{c}_{ab}= \Gamma_{ab}^{c}+ \Gamma_{ab}^{i}A_{i}^{c}
\nonumber\\
&\g^{j}_{ai}= \Gamma_{ai}^{j}- \Gamma_{ab}^{j}A_{i}^{b}
\nonumber\\
&\g^{j}_{ia}= \Gamma_{ia}^{j}- \Gamma_{ba}^{j}A_{i}^{b}
\nonumber\\
&\g^{b}_{ia}= \Gamma_{ia}^{b}- \Gamma_{ca}^{b}A_{i}^{c} 
+\Gamma_{ia}^{j}A_{j}^{b}-\Gamma_{ca}^{j}A_{i}^{c}A_{j}^{b}
\nonumber\\
&\g^{b}_{ai}= \Gamma_{ai}^{b}- \Gamma_{ac}^{b}A_{i}^{c} 
+\Gamma_{ai}^{j}A_{j}^{b}-\Gamma_{ac}^{j}A_{i}^{c}A_{j}^{b}
\nonumber\\
&\g^{k}_{ij}= \Gamma_{ij}^{k}- \Gamma_{aj}^{k}A_{i}^{a} 
-\Gamma_{ia}^{k}A_{j}^{a}+\Gamma_{ab}^{k}A_{i}^{a}A_{j}^{b}
\nonumber\\
&\g^{a}_{ij}= \Gamma_{ij}^{a}- \Gamma_{bj}^{a}A_{i}^{b} 
-\Gamma_{ib}^{a}A_{j}^{b}+\Gamma_{bc}^{a}A_{i}^{b}A_{j}^{c}
-\Gamma_{bj}^{k}A_{i}^{b}A_{k}^{a}-
\Gamma_{ib}^{k}A_{j}^{b}A_{k}^{a}
+\Gamma_{bc}^{k}A_{i}^{b}A_{j}^{c}A_{k}^{a}\nonumber\\
&-\d_{i}A_{j}^{a} + \Gamma_{ij}^{k}A_{k}^{a}
\eea

Paradoxical though it may be seen the 
task to calculate the reduced
Levi-Civita 
connection is shorter than the usual one, because the 
knowledge of
its invariance
under the adapted  diffeomorphismes (\ref{ad}) 
drops out the terms
proportiponally to 
$A_{i}^{a}$ being non derivatives 
(we can call it $A=0$ projection).
Therefore (\ref{1100}) must 
be understood as $\g\ic = ( \Gamma\ic -
\d_{\mu}J_{\nu}^{\rho})\vert_{A=0}$. It is just 
the need to know $\Gamma\vert_{A=0}$ instead of  
$\Gamma$ itself ,
which makes easier to 
get the reduced version.

The usual curvature, torsion and Ricci tensor's 
expression for a generic
connection 
$\Gamma_{\mu\nu}^{\rho}$ are :

\bea
\label{12200}
&R(\Gamma)_{\mu\nu\sigma}^{\rho}=
\d_{\mu}\Gamma_{\nu\sigma}^{\rho}
- \d_{\nu}\Gamma_{\mu\sigma}^{\rho}-
 \Gamma_{\mu\sigma}^{\beta}\Gamma_{\nu\beta}^{\rho} 
+ \Gamma_{\nu\sigma}^{\beta}\Gamma_{\mu\beta}^{\rho} 
\nonumber\\
&T(\Gamma)_{\mu\nu}^{\rho}={1\over2}(\Gamma_{\mu\nu}^{\rho}
-\Gamma_{\nu\mu}^{\rho})\nonumber\\
&R(\Gamma)_{\mu\nu}= R(\Gamma)_{\mu\rho\nu}^{\rho}
\eea

The reduced Riemaniann curvature for the $U(1)^n$ is given by 

\bea
\label{1200}
&r_{abcd}= {1\over4}(\hat{\d}^{i}G_{ac}\d_{i}G_{bd} -
\hat{\d}^{i}G_{bc}\d_{i}G_{ad})
\nonumber\\
&r_{abci}= {1\over4}F_{ij}^{e}
\hat{\d}^{j}G_{cd}(G_{ea}\delta_{b}^{d} - 
G_{eb}\delta_{a}^{d})\nonumber\\
&r_{ijab}={1\over4}(G^{cd}\d_{i}G_{ac}\d_{j}G_{bd} + 
G_{ac}G_{bd}\hat{F}_{i}^{c\,k}F_{jk}^{d} - (i\,j
\longleftrightarrow j\,i) )\nonumber\\
&r_{aibj}={1\over2}\hat{\D}_{i}\hat{\D}_{j}G_{ab}-{1\over4}
G^{cd}\d_{i}G_{bd}\d_{j}G_{ac}-{1\over4}
G_{ac}G_{bd}\hat{F}_{i}^{d\,k}F_{jk}^{c}
\nonumber\\
&r_{akij}=-{1\over2}\hat{\D}_{k}(G_{ab}F_{ij}^{b})+
{1\over4}(F_{ki}^{b}\d_{j}G_{ab} - F_{kj}^{b}\d_{i}G_{ab})
\nonumber\\
&r_{ijkl}=\hat{R}_{ilkl}+
{1\over4}G_{ab}(F_{ik}^{a}F_{jl}^{b}-F_{jk}^{a}F_{il}^{b}
+2F_{ij}^{a}F_{kl}^{b})
\eea

where the hatted objects are the ones calculated 
with the quotient metric
$\bar{G}_{ij}$, 
$F^{a}_{ij}\equiv \d_{i}A^{a}_{j} - \d_{j}A^{a}_{i}$, and
$A_{i}^{a}=G^{ab}A_{ib}$.
The Riemaniann curvature is obtained using 
(\ref{rt})(\ref{j})(\ref{j2}), 
i.e. , $R=J(-A)r$.
\section{Apendix :The generalized curvature}
In this section I will  write  the reduced generalized curvature
needed for the fifth section's calculations.
To do it in a T-duality 
suitable way means the introduction of the 
$\hat{D}_{i}^{\pm}\Theta\equiv
\hat{\D}_{i}^{\pm}+ {\Delta_{\Theta}\over2}\d_{i}\ln\G)
\Theta$ T-duality covariant derivative. As in (\ref{tdcd}) 
$\Delta_{\Theta}$ is the $\Theta$'s T-duality weight.

In terms of $\G$ and $F_{ij}^{\pm}\equiv \g_{ij}^{\pm\,0}$ the 
relevant components are:

\bea
\label{grcxx}
 &r^{\pm}_{0i0j} =\{{1\over2}\hat{D}_{i}^{\pm}\d_{j}\G
+\G^2\hat{F}_{i}^{\pm\,k}
F_{kj}^{\mp}\}+{\G\over4}\d_{i}\ln\G\d_{j}\ln\G
 \nonumber\\
&r^{\pm}_{0ijk} =\{\G\hat{D}_{i}^{\pm}F_{jk}^{\mp}+{\G\over2}
(F_{ik}^{\pm}\d_{j}\ln\G - F_{ij}^{\pm}\d_{k}\ln\G)\}
+{\G\over2}F_{jk}^{\mp}\d_{i}\ln\G
 \nonumber\\
&r^{\pm}_{ijkl} =\{\hat{R}_{ijkl}^{\pm}+
\G(F_{ik}^{\pm}F_{jl}^{\pm}
 - F_{jk}^{\pm}F_{il}^{\pm})\}+\G(\g_{ij}^{\pm}+
\g_{ij}^{\mp})\g_{kl}^{\mp}
\eea

where $\hat{R}_{ijkl}^{\pm}= R(\hat{\Gamma} \pm 
\hat{h})_{ijkl}$.
The other components are related by the symmetry properties 
$ R^{\pm}_{\mu\nu\sigma\rho}= -
R^{\pm}_{\nu\mu\sigma\rho}
= - R^{\pm}_{\mu\nu\rho\sigma}$ and 
$ R^{\pm}_{\mu\nu\rho\sigma}=
R^{\mp}_{\rho\sigma \mu\nu}$.
The terms in brackects are the ones 
transforming with the dominant
$(-1)^{(g_{\mu}+g_{\nu})}
(\pm\G)^{-n_{0}}$,
while the remaining terms transform with the 
$-(-1)^{(g_{\mu}+g_{\nu})}
(\pm\G)^{-n_{0}}$ giving the inhomogeneus part $-{2\over K^2}
\D_{\mu}^{\pm}K_{\nu}\D_{\sigma}^{\mp}K_{\rho}$.

\end{document}